# Artificial neural networks
# for nonlinear pulse shaping in optical fibers

**Sonia Boscolo [1] and Christophe Finot [2,*]**

*[1] Aston Institute of Photonic Technologies, School of Engineering and Applied Science, Aston University, Birmingham B4 7ET, United Kingdom*

*[2] Laboratoire Interdisciplinaire Carnot de Bourgogne, UMR 6303 CNRS-Université de Bourgogne-Franche-Comté, 9 avenue Alain Savary, BP 47870, 21078 Dijon Cedex, France*

*christophe.finot@u-bourgogne.fr*

*Tel.: +33 3 80395926*

**Abstract:** We use a supervised machine-learning model based on a neural network to predict the temporal and spectral intensity profiles of the pulses that form upon nonlinear propagation in optical fibers with both normal and anomalous second-order dispersion. We also show that the model is able to retrieve the parameters of the nonlinear propagation from the pulses observed at the output of the fiber. Various initial pulse shapes as well as initially chirped pulses are investigated.

**Keywords:** pulse shaping, neural networks, nonlinear propagation.

# I.     Introduction

Pulse shaping based on nonlinear phenomena in optical fibers has become a remarkable tool to tailor the spectral and temporal content of light signals [1, 2], leading to the generation of a large variety of optical waveforms such as ultra-short compressed pulses [3], parabolic- [4], triangular- [5] and rectangular- [6] profiled pulses. Very different features of the nonlinear pulse evolution can be observed depending on the sign of the fiber group-velocity dispersion (GVD), which result in specific changes of the pulse temporal shape, spectrum and phase profile. Specifically, because the nonlinear dynamics of pulses propagating in fibers with normal GVD are generally sensitive to the initial pulse conditions, it is possible to nonlinearly shape conventional laser pulses into various specialized waveforms through control of the initial pulse temporal intensity and/or phase profile [1]. Conversely, the temporal and spectral features of pulses propagating in fibers with anomalous GVD are typically governed by soliton dynamics [7]. Yet, due to the typically wide range of degrees of freedom involved, predicting the behavior of nonlinear pulse shaping can be computationally demanding, especially when dealing with inverse problems.

Owing to its power of extracting essential information from large amounts of data, machine learning is bringing a revolutionary reform to research in the physical sciences [8]. In the field of photonics, a number of studies have been recently reported in laser design and optimization [9-11], complex nonlinear dynamics [12], design of photonic crystal fibers and optical components [13, 14], pulse characterization [15], and optical communications [16, 17]. In [18], we have shown that the combination of a graphical approach with the machine-learning method of neural networks (NNs) can provide a rapid and precise identification of the parameters of nonlinear pulse shaping systems based on pulse propagation in a normally dispersive fiber that are required to generate pulses with preset temporal features. In this paper, we use a supervised learning model based on a NN to predict the temporal and spectral intensity profiles of the pulses that form upon nonlinear propagation in fibers with both normal and anomalous dispersion. We also assess the ability of the NN to solve the inverse problem of determining the nonlinear propagation properties from the pulses observed at the fiber output and to classify the output pulses according to the initial pulse shape. Furthermore, we show how our model can handle the nonlinear shaping of initially chirped pulses.

## II.      Problem under study and numerical tools

The general scheme for nonlinear shaping that we consider in this paper comprises a pre-chirping stage followed by a nonlinear propagation stage. Within this scheme, an initial pulse $\psi_0(t)$ with a peak power $P_0$ is first propagated through a dispersive medium, such as a pair of diffraction gratings, a prism pair [19], a segment of hollow core or standard fiber with very low nonlinearity [20, 21]. This linear propagation imprints a parabolic spectral phase onto the pulse: $C_0 \, \omega^2 / 2$, where the chirp coefficient $C_0$ equals the cumulative GVD of the medium ($\omega$ being the angular frequency). The so obtained chirped pulse is then propagated through a fiber that reshapes both its temporal and spectral intensity profiles. Our main interest here is in these physical quantities of the pulse that can be directly recorded in experiments, rather than in the complex envelope of the electric field. Different initial pulse shapes are studied: a Gaussian pulse $\psi_0(t) = \sqrt{P_0} \exp\left(-t^2 / 2T_0^2\right)$, a hyperbolic secant pulse $\psi_0(t) = \sqrt{P_0} \operatorname{sech}\left(t / T_0\right)$, a parabolic pulse $\psi_0(t) = \sqrt{P_0} \sqrt{1 - t^2 / T_0^2} \; \theta\left(T_0 - |t|\right)$ and a second-order super-Gaussian pulse $\psi_0(t) = \sqrt{P_0} \exp\left(-t^4 / 2T_0^4\right)$. Here, $T_0$ is a characteristic temporal value of the pulse and $\theta(x)$ is the Heaviside function.

To generate data that characterizes the nonlinear shaping process, we use the nonlinear Schrödinger equation (NLSE) [7], which can describe well a variety of nonlinear phenomena associated with pulse propagation in fibers in spite of the fact that it only includes two physical effects, namely, linear GVD and nonlinear self-phase modulation (SPM). The NLSE for the complex electric field envelope, $\psi(z,t)$, is written as

$$i \frac{\partial \psi}{\partial z} - \frac{\beta_2}{2} \frac{\partial^2 \psi}{\partial t^2} + \gamma \, |\psi|^2 \, \psi = 0 \qquad (1)$$

where $z$ is the propagation coordinate, $t$ is the retarded time, and $\beta_2$ and $\gamma$ are the respective GVD and Kerr nonlinearity coefficients of the fiber. Note that the effects of linear loss can be neglected given the very low loss of silica fibers in the telecommunication wavelength window. Here we also neglect higher-order linear and nonlinear effects as the leading-order behavior is well approximated by Eq. (1). It is useful to normalize Eq. (1) by introducing the dimensionless variables: $u = \psi / \sqrt{P_0}$, $\xi = z/L_D$, $\tau = t/T_0$, and write it in the form

$$i \frac{\partial u}{\partial \xi} - \frac{\mathrm{sgn}(\beta_2)}{2} \frac{\partial^2 u}{\partial \tau^2} + N^2 |u|^2 u = 0 \qquad (2)$$

where $L_D = T_0^2/|\beta_2|$ and $L_{NL} = 1/(\gamma P_0)$ are the respective dispersion length and nonlinear length associated with the initial pulse, and the parameter $N$ ('soliton-order' number) is introduced as $N^2 = L_D/L_{NL}$. This way, the nonlinear shaping problem, which depends on the six physical parameters ($T_0$, $P_0$, $C_0$, $\beta_2$, $\gamma$, $L$) where $L$ is the fiber length, is mapped onto a problem in the three-dimensional space of $(\xi, N, C = C_0/T_0^2)$. This dimensionality reduction relaxes the complexity of the problem, and for a specific selected set of normalized parameters there are many groups of physical parameters suitable the defining equations of $\xi$, $N$ and $C$.

Equation (2) is solved with a standard split-step Fourier propagation algorithm [7], using a uniform grid of $2^{13}$ points on a time window of length 80 $T_0$. In order to generate the temporal and spectral properties for the NN, we perform anamorphic sampling of the output intensity profile provided by Eq. (2). This enables us to well represent both the details of the short pulses that can be encountered in the anomalous dispersion regime of the fiber and the longer pulses that are observed in the normal dispersion regime. Given the symmetry of the problem (symmetry of the NLSE and the initial temporal condition), we can restrict our sampling to the positive times and frequencies only. This way we can maintain the number of useful points moderate. Hence, we consider $A = 65$ points on the interval 0 to 25 $T_0$ to represent the temporal intensity profiles of the pulses, and $B = 35$ points on the interval 0 to $3.4/T_0$ for the spectral intensities. The sampled data and the initial conditions are used to train a NN and validate its predictions. We employ a feed-forward NN relying on the Bayesian regularization back propagation algorithm and including three hidden layers with fourteen neurons each, as shown in Fig. 1. This NN is programmed in Matlab using the neural network toolbox.

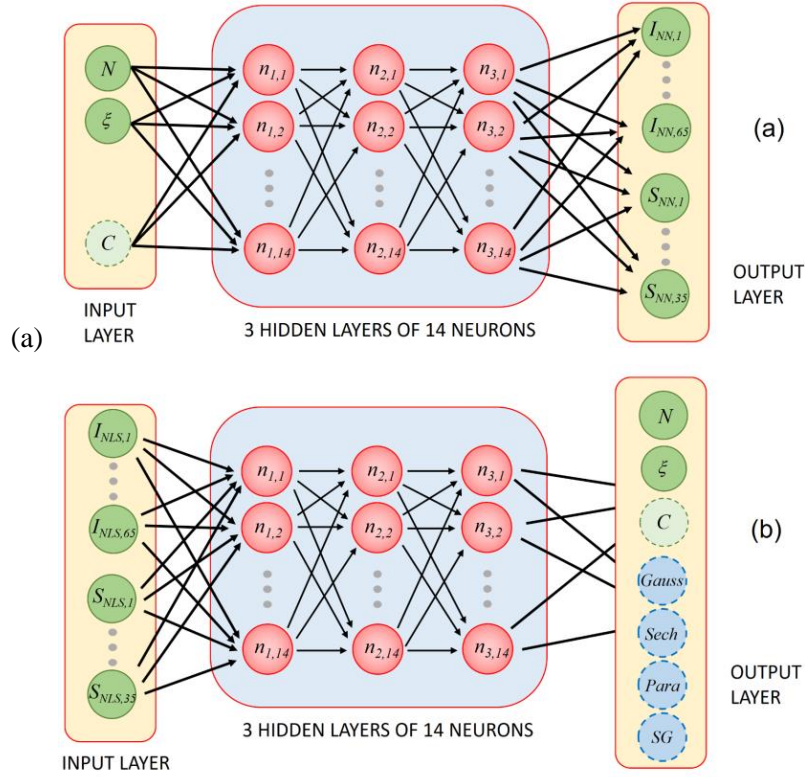

Figure 1: Neural network model used to: (a) predict the output pulse intensity profiles from the fiber, and (b) to retrieve the nonlinear propagation properties.

## III. Nonlinear shaping of initially transform-limited pulses

In this section, we study the problem of the nonlinear shaping of initially-transform limited pulses (i.e., with $C_0 = 0$), which can be handled in the parameter space of normalized propagation length and soliton-order number, $(\xi, N)$.

### A/ Prediction of the output pulse properties

The training step here involves feeding the NN with an ensemble of 4500 data (both temporal and spectral) generated from numerical simulations of the NLSE and associated with the propagation of a Gaussian pulse with $\xi$ ranging from 0.025 to 2.5. 1500 randomly chosen data points are used to cover the parameter space in the normal dispersion regime of the fiber with $N$ ranging from 0.025 to 4, and 3000 randomly chosen points for the anomalous dispersion regime with $N$ ranging from 0.025 to 3. After training, the NN is tested on a distinct ensemble of 25000 data not used in the training step. Figure 2 shows the temporal and spectral profiles of the pulse obtained from the network for $\xi = 2$ and $N = 4$ when the fiber has normal GVD, and for $\xi = 1.8$ and $N = 2.5$ at anomalous dispersion. The predictions from the NN algorithm show excellent agreement with the results of the NLSE propagation model over most part of the pulse shape and spectrum. The network is able to reproduce the large temporal and spectral broadening experienced by the pulse upon propagation in the normal dispersion regime. With anamorphic sampling of its output, the network is also able to resolve the details of the temporally compressed pulse and the concomitant more complex structure of the spectrum that are observed after propagation at anomalous dispersion. Nevertheless, some discrepancies with the expected results are visible, and are more pronounced in the anomalous regime in which the propagation dynamics are more complex.

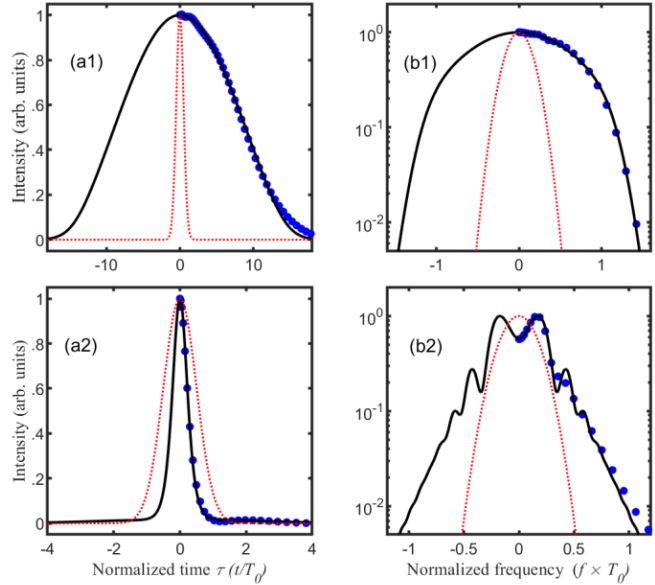

Figure 2: Temporal (panels 1) and spectral (panels 2) intensity profiles of an initial Gaussian pulse after propagation in (a) a normally dispersive fiber with $\xi = 2$ and $N = 4$, and (b) an anomalously dispersive fiber with $\xi = 1.8$ and $N = 2.5$. The predictions from the NN algorithm (blue circles) are compared with the results of NLSE numerical simulations (black curves). Also shown are the input intensity profiles (red dotted curves

As a measure of the prediction error of the NN algorithm we use the parameter of misfit between the (normalized) output temporal or spectral intensity profile generated by the network, $I_{NN}/S_{NN}$, and the expected profile produced by numerical simulation of the NLSE, $I_{NLS}/S_{NLS}$:

$$M_T = \sum_{i=1}^{A} \left( I_{NN,i} - I_{NLS,i} \right)^2 \Big/ \sum_{i=1}^{M} I_{NLS,i}^2$$
$$M_F = \sum_{i=1}^{B} \left( S_{NN,i} - S_{NLS,i} \right)^2 \Big/ \sum_{i=1}^{M} S_{NLS,i}^2$$

(3)

In Eq. (3), the expected profiles are interpolated to the same time or frequency points used for sampling the network output. The results obtained for 25000 randomly chosen combinations of input parameters $\xi$ and $N$ in the normal and anomalous dispersion regimes of the fiber are summarized in Fig. 3(a), and confirm that the nonlinear pulse shaping occurring in a fiber with normal GVD is easier to predict than that occurring in the anomalous dispersion regime. This can be physically explained by the more complex nonlinear pulse dynamics that take place in the presence of anomalous dispersion, involving compression and/or splitting stages over short propagation distances. The distributions of values of the temporal and spectral misfit parameters (Fig. 3(b)) show that, remarkably, more than 90% of the error realizations are well confined to values below 0.04, but some error values are spread out over a wider range. This deviation mostly occurs in the anomalous dispersion regime.

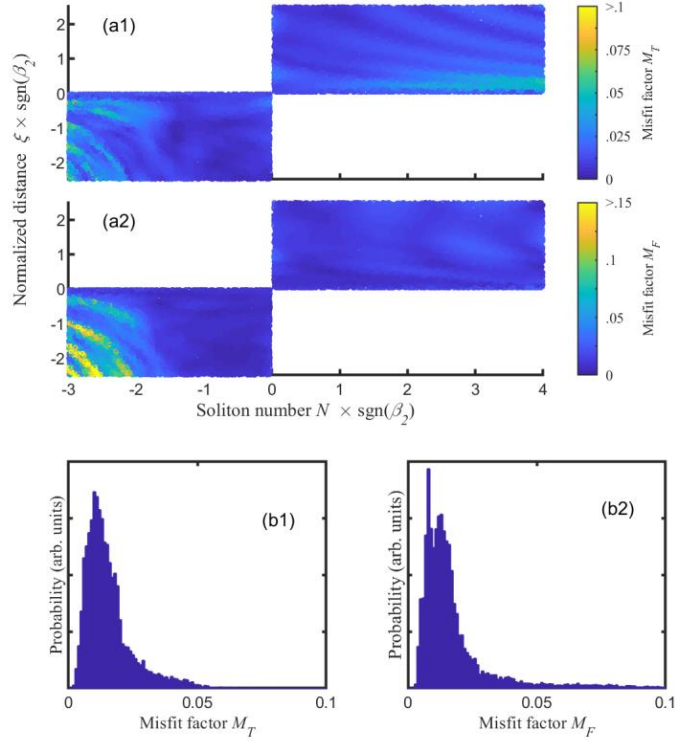

Figure 3: (a) Maps of misfit parameter values between the NN predictions and the NLSE simulation results for the output pulse shape (subplot a1) and optical spectrum (subplot a2) in the two-dimensional space $(N, \xi)$, relating to the propagation of a Gaussian pulse in the normal and anomalous dispersion regimes of the fiber with randomly chosen combinations of $\xi$ and $N$. (b) Distribution densities of the temporal (subplot b1) and spectral (subplot b2) misfit parameter values.

## B/ Retrieval of the propagation characteristics

The inverse problem at hand is much more complex: from a pulse shape and spectrum that are generated after propagation in the fiber, the properly trained network should be able to retrieve the parameters $\xi$, $N$ as well as the regime of dispersion of the fiber. One could tackle this inverse problem by using reverse propagation, that is, solving the NLSE to the backward direction [22, 23]. However, this method would require knowledge of the peak power (or energy) of the pulse at the fiber output, whereas we only consider here the pulse shape characteristics.

For the inverse NN training phase, we use the same data set as that used for the direct problem. The results of testing the trained NN on 25000 randomly chosen new simulated output pulses are shown in Fig. 4. We can see from Fig. 4(a) that the estimation error on $\xi$ ($\Delta\xi$) and $N$ ($\Delta N$) (defined as the difference between the retrieved parameter value and the target value extracted from the NLSE simulation data) is very close to zero for all test realizations except those corresponding to low input powers or short propagation lengths for which the changes in the temporal and spectral shapes of the propagating pulse are negligibly small, thus leading to similar shapes for different $\xi$ values. If we limit the data points to the ranges $\xi \geq 0.25$ and $N \geq 0.25$, then, interestingly, each temporal and spectral shape can be unambiguously associated with a single parameter set ($\xi$, $N$). The network is able to work out the sign of the fiber dispersion perfectly. The distribution densities of the estimation errors shown in Fig. 4(b) confirm the remarkable accuracy of the results obtained with the NN algorithm: the root-mean square errors on $\xi$ and $N$ are below $2 \ 10^{-3}$ and $2.7 \ 10^{-3}$, respectively.

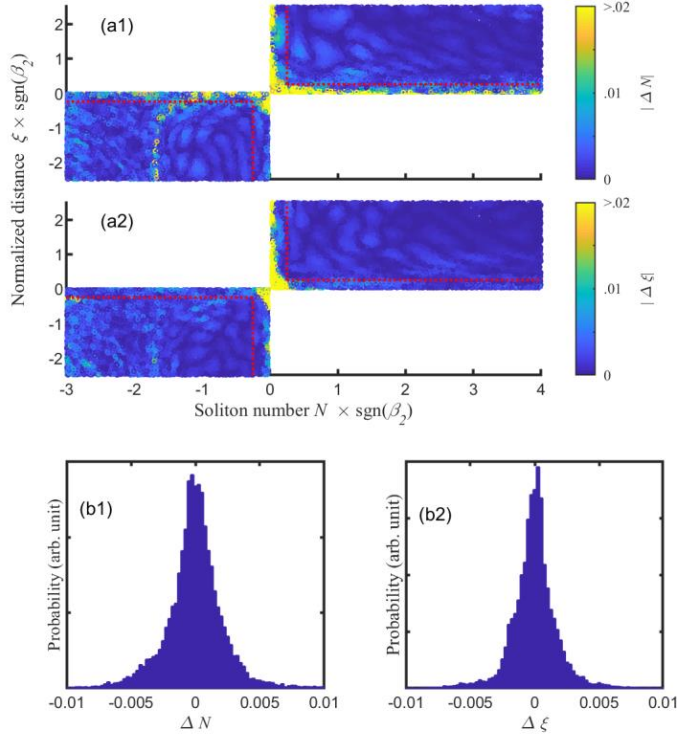

Figure 4: (a) Maps of estimation error values on the soliton number $N$ (subplot a1) and normalized propagation length $\xi$ (subplot a2) in the two-dimensional space ($N$, $\xi$) for both normal and anomalous dispersion, when the NN is interrogated with randomly chosen new simulated output pulses from the fiber. The red dashed lines delimit the data domain that is used for the statistical error analysis. (b) Distribution densities of the estimation errors on $N$ (subplot b1) and $\xi$ (subplot b2).

We also study the effect of the initial pulse shape on the network's ability to solve the inverse problem. Maps of values of the estimation error on the normalized propagation length are provided in Fig. 5 for 25000 test realizations for each of the initial shapes: hyperbolic secant, parabolic and super-Gaussian pulse. A difficulty arises with the hyperbolic secant pulse propagating in the anomalous regime of the fiber: the estimation error is non-negligible for values of the soliton-order number around 1. This originates in the evolution of the fundamental soliton ($N$=1), which propagates without changing its shape, thereby making the propagation length difficult to evaluate. Furthermore, higher-order solitons (characterized by integers $N$>1) follow a periodic evolution pattern, which entails that the same pulse shape may be associated with different propagation lengths [24]. We expect that the effects of energy dissipation or higher-order linear /nonlinear propagation effects, can alleviate this ambiguity.. Initially parabolic pulses can be better handled by the inverse problem NN. The model still works with initially super-Gaussian pulses, whose dynamics can be rather different especially in the anomalous dispersion regime in which the waveforms undergo splitting followed by compression [25]. We can also see from Fig. 5 that, again, the anomalous dispersion regime of the fiber is more complex to deal with.

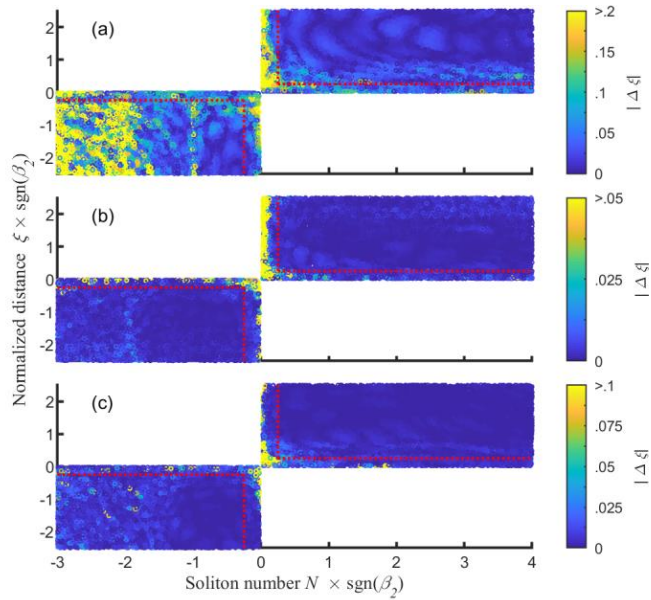

Figure 5: Maps of estimation error values on the normalized propagation length $\xi$ in the two-dimensional space ($N$, $\xi$) for both normal and anomalous dispersion, when the NN is interrogated with randomly chosen new simulated output pulses from the fiber corresponding to input (a) hyperbolic secant, (b) parabolic and (c) super-Gaussian pulses.

## C/ Identification of the initial pulse shape

For this problem, we train the network on an ensemble of 16000 simulated output pulses from the fiber corresponding to a mix of Gaussian, hyperbolic secant, parabolic and super-Gaussian initial pulse shapes and randomly chosen combinations of input parameters $\xi$ and $N$. The data relating to the propagation of hyperbolic secant pulses in the anomalous dispersion regime is excluded from the training set.. Then we ask the trained network to categorize $10^5$ new unlabeled simulated output pulses (distinct from the training data) according to the initial waveform and to retrieve the associated propagation parameters. As we can observe from Fig. 6(a), the classification accuracy of the NN algorithm is remarkably high: there are only 9 errors, which represent less than .01% of the total number of input samples. The estimation error on the propagation length has increased compared to the inverse problem studied in the previous section, where a single initial waveform is considered at a time. After exclusion of the data points falling into the critical parameter region ($\xi$ or $N$ below 0.25), we can expect rms errors on the propagation length and soliton-order number of 0.076 and 0.085, respectively [Fig. 6(b)].

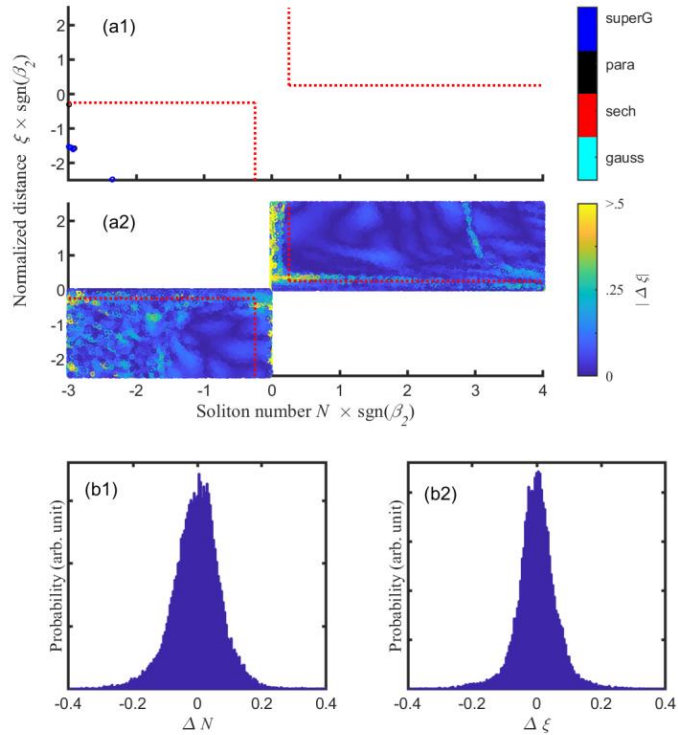

Figure 6: Points where errors in the detection of the initial pulse shape occur (subplot a1) and map of estimation error values on the normalized propagation length $\xi$ (subplot a2) in the two-dimensional space ($N$, $\xi$) for both normal and anomalous dispersion, when the NN is interrogated with randomly chosen new simulated output pulses from the fiber corresponding to an unlabeled mix of input Gaussian, hyperbolic secant, parabolic and super-Gaussian pulses. In subplot a1, the color of the circle indicates the class label that is not predicted correctly. The red dashed lines delimit the data domain that is used for the statistical error analysis. (b) Distribution densities of the estimation errors on $N$ (subplot b1) and $\xi$ (subplot b2).

# III.    Nonlinear shaping of initially chirped pulses

In this section, we include an additional degree of freedom in our analysis by considering a possible pre-chirping of the pulses that are transmitted through the fiber. Our focus is here on pulse propagation in the normal regime of dispersion. Depending on the chirp of the input pulse, the initial stage of nonlinear dynamics in the fiber, where Kerr-induced SPM dominates over GVD, may be very different. Indeed, input pulses with a negative chirp coefficient $C_0$ will experience spectral compression as a result of SPM [26, 27], whereas for initially positively chirped (or Fourier transform-limited) pulses, spectral broadening will drive the nonlinear dynamics and eventually lead to optical wave-breaking [28]. In [18], we used a graphical method to find the combinations of values for the parameters $\xi$, $N$ and $C$ that support the formation of pulses with specified temporal features in the fiber.

## A/ Prediction of the output pulse properties

The NN learns the NLSE model from an ensemble of $2 \ 10^4$ simulations of the propagation of an initial Gaussian pulse with randomly chosen combinations of values for $\xi$, $N$ and $C$ over the ranges 0.025 to 3, 1.5 to 5, and -3.4 to 3.4, respectively. For this configuration, the data is sampled over a temporal interval of 30 $T_0$ and a spectral window of $2.3/T_0$. We then test the trained NN on $7 \ 10^4$ simulations from a distinct ensemble of random initial conditions. Figure 7 shows typical examples of the nonlinear pulse shaping occurring in the fiber: formation of parabolic-like pulses with rectangular-like spectra when the chirp of the input pulses is positive [subplot (a)] [29], and narrowing of the pulse spectrum for negative input chirp [subplot (b)] [27]. It is clear that the NN algorithm performs impressively in reproducing the output pulse shapes.

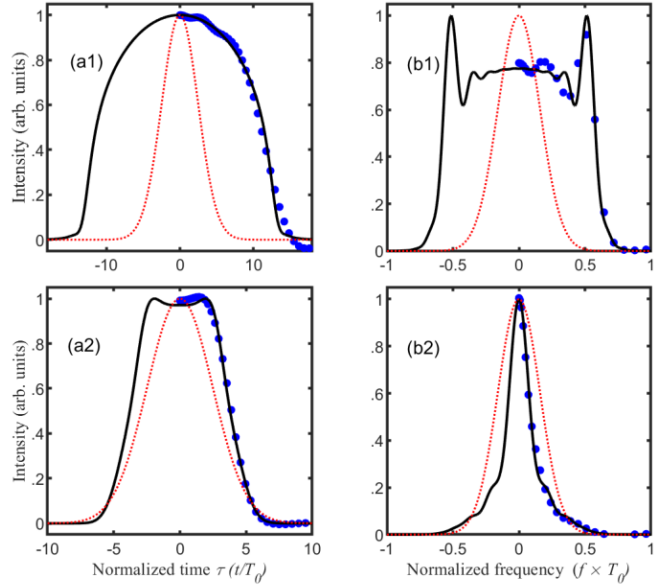

Figure 7: Temporal (panels 1) and spectral (panels 2) intensity profiles of an initial Gaussian pulse after propagation in a normally dispersive fiber with (a) $\xi = 2.3$, $N = 3$, and $C = -2.42$, and (b) $\xi = 2$, $N = 3$ and $C = 2.42$. The predictions from the NN algorithm (blue circles) are compared with the results of NLSE numerical simulations (black curves). Also shown are the input intensity profiles (red dotted curves).

## B/ Inverse problem

The results of testing the inverse problem NN on $7 \ 10^4$ new simulated output pulses are summarized in Fig. 8. We note that the fiber nonlinearity is essential to distinguishing the combinations of $\xi$, $N$ and $C$ values that are associated with each test pulse. Indeed, pulse pre-chirping entails a drop of the pulse peak power, which in turn results in a frequent occurrence of estimation errors on $\xi$, $N$ and $C$ for the test realizations that relate to low values of $N$. Therefore, we restrict the statistical error analysis to the test data associated with $N > 1.5$. We can see from Fig. 8 that the values of the input parameters obtained from the network algorithm are in agreement with the known values from the simulation data, but the estimation errors are higher than those made in the case of initially transform-limited pulses: the rms deviations are 0.36, 0.09 and 0.55 for $\xi$, $N$ and $C$ respectively.

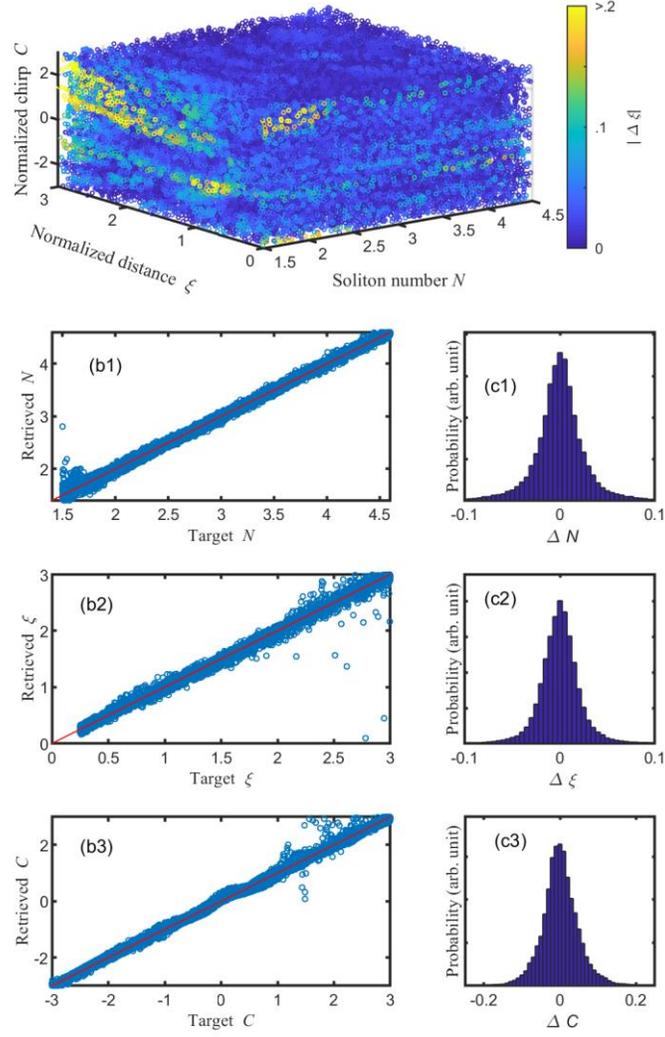

Figure 8: (a) Map of estimation error values on the normalized propagation length $\xi$ in the three-dimensional space ($N$, $\xi$, $C$), when the NN is interrogated with randomly chosen simulated output pulses from the fiber. (b) Regressions between the predicted values of $N$, $\xi$ and $C$ from the NN algorithm and the exact target values from the simulated data. (c) Distribution densities of the estimation errors on $N$, $\xi$ and $C$.

# IV.    Conclusion

We have successfully used a supervised machine-learning model based on a NN to solve both the direct and inverse problems relating to the shaping of optical pulses that occur upon nonlinear propagation in optical fibers. Remarkably, within the range of system parameters considered, any temporal and spectral shape generated at the fiber output from the propagation of an initially Fourier-transform limited Gaussian pulse can be unambiguously associated with a single set of normalized propagation length and soliton-order number values, $(\xi, N)$. Our results show that a properly trained network can greatly help the design and characterization of fiber-based shaping systems by providing immediate and sufficiently accurate solutions. We have limited our discussion here to passive non-dissipative propagation. An interesting extension of the method described will be to the analysis of the attraction of parabolic pulses towards a self-similar state in normally dispersive nonlinear fibers with linear gain [4]. Furthermore, although demonstrated here in a fiber optics context, the principle of using NN architectures to solve wave equation-based inverse problems is expected to apply to many physical systems [30, 31].

## Acknowledgements

We acknowledge the support of the Institut Universitaire de France (IUF).